\documentclass[prb,aps,twocolumn,floats,epsfig]{revtex4}
\usepackage{amssymb}
\usepackage{amsbsy}
\usepackage{amsmath}
\usepackage{wasysym}
\usepackage{bm}
\usepackage{epsfig}

\begin{document}

\title{Manipulating unpaired Majorana fermions in a quantum spin chain}

\author{Abhinav Saket, S. R. Hassan and R. Shankar}

\affiliation{The Institute of Mathematical Sciences, C.I.T. Campus, 
Chennai 600 113, India}

\date{\today}

\begin{abstract}

We analyse an exactly solvable spin-$1/2$ chain which is a generalised version
of Kitaev's honeycomb model. We show that every state of the system has a
$2^{N/4}$ fold degeneracy, where $N$ is the number of sites. We present
analytic solutions for the zero energy modes of the Majorana fermions.
Localised, unpaired Majorana modes occur even in the bulk of the chain and they
are bound to kink (anti-kink) $Z_2$ flux configurations. The unpaired Majorana
modes can therefore be created and manipiulated if the $Z_2$ flux
configurations can be controlled. We delineate the regions in parameter space
for homogenous chains where the zero modes occur. We further show that there is
a large parameter space for inhomogenous chains where the unpaired modes occur
and that their wavefunctions can be tuned if the couplings of the model can be
tuned.

\end{abstract}


\maketitle

\section{Introduction}
                              
The idea of topological quantum computation as a way to incorporate fault 
tolerence at the hardware level has been getting a lot of attention recently
\cite{tqc}. In this scheme, qubits are non-abelian anyons and the braiding 
operations on them implement quantum gates. One of the simplest class of 
non-abelian anyons are realised in systems with unpaired Majorana fermions
(UMF) \cite{umf}. In a fermionic system with $N$ zero energy modes, there 
are $2N$ Majorana modes. If these $2N$ modes can be independently moved around 
each other, then the geometric phases picked up correspond to a non-abelian 
representation of the braid group. Thus it is of interest to study physically
realisable systems where UMF exist and can be manipulated.

Non-abelian anyons are theoretically predicted to occur in certain fractional
quantum Hall states like $\nu=5/2$ \cite{mooreread}. There is also theoretical
work showing how they could be realised in quantum circuits \cite{faroaetal}.
In this context, Kitaev presented a remarkable solvable spin-1/2 model 
on a honeycomb lattice \cite{kitaevhc} which realises non-abelian anyons made 
up of UMF. The model can be written in terms of Majorana
fermions in the background of $Z_2$ gauge fields. This fermionisation procedure
is very similar to that proposed in the context of the resonating valence
bond (RVB) theory \cite{gbpwa}. The crucial difference is that in Kitaev's 
honeycomb model, the gauge fields are constants of motion, hence the RVB type 
mean field theory is exact and the problem reduces to solving a theory of 
non-interacting Majorana fermions in the background of static $Z_2$ gauge field
configurations. Kitaev showed that the ground state is the flux free 
configuration. The model has a phase which is characterised by a topological 
invariant, the Chern number, being equal to $\pm 1$. In this phase there are 
UMF trapped to each vortex (a $Z_2$ flux). Thus if vortices can be created and
manipulated, it is possible to braid the UMF. However, the $Z_2$ flux operator
is a 6-spin operator and thus may not be easy to realise in practice.

Kitaev's honeycomb model can be generalised to a variety of other lattices.
It can be constructed on any lattice with coordination number three, if all the 
bonds can be coloured using three colours. It has been shown that all such 
models can be realised in cold atom systems \cite{demler} and using quantum 
circuits \cite{nori}. 

In paper we present a one dimensional generalisation of Kitaev's honeycomb 
model which we call the tetrahedral chain (TC) and analyse its zero energy 
modes in detail. One dimensional models with UMF at the edges have been 
studied earlier \cite{1dkit,1dumf}. The new feature of the TC is that the 
wavefunctions of the UMF not necessarily peaked at the edges of the chain but
can be peaked anywhere in the bulk. As we will show, they are trapped to 
kink and anti-kink flux configurations and can be moved by tuning the 
flux configuration. Further, by tuning the coupling constants, their 
wavefunctions, which we obtain analytically, can also be tuned. 

The rest of this paper is organised as follows. In section \ref{tmod} we 
present the model and its conserved quantities. Section \ref{jwt} describes 
the Jordan-Wigner transformation which enables us to rewrite the theory in 
terms of Majorana fermions hopping in the background of static $Z_2$ gauge 
field configurations. The diagonalisation of the Majorana fermion problem is 
described in section \ref{diag} and the numerical calculations to determine
the ground state flux configurations are presented in section \ref{num}.
Section \ref{degen} gives an analytic proof of the degeneracy of the 
eigenstates that we find numerically. The dependence of the gap as a function
of the couplings is computed in section \ref{gs} and a gapless line in 
the parameter space is identified. Section \ref{zmk} contains a detailed 
analysis of the zero-modes of the Majorana fermions. We summarise our results
and discuss them in the conluding section \ref{concl}.

\section{The Tetrahedral Model}
\label{tmod}

\begin{figure}[h] 
\includegraphics[width=\linewidth]{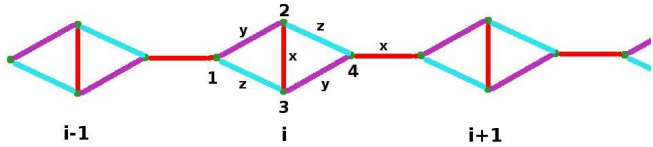}
\caption{The tetrahedral Kitaev chain. There are four sites per unit
cell. The $x,y$ and $z$ bonds are as indicated.}
\label{thkc1}
\end{figure}

The chain we define our model on is shown in Fig. \ref{thkc1}. There
are four sites per unit cell which are labelled as shown in the figure.
The hamiltonian is,
\begin{eqnarray}
\nonumber
H&=&\sum_i\left( 
J_x\left(\sigma^x_{i-1,4}\sigma^x_{i,1}+\sigma^x_{i,2}\sigma^x_{i,3}\right)
\right.\\
\nonumber
&&+J_y\left(\sigma^y_{i,1}\sigma^y_{i,2}+\sigma^y_{i,3}\sigma^y_{i,4}\right)
\\
\label{ham1}
&&\left.
+J_z\left(\sigma^z_{i,1}\sigma^z_{i,3}+\sigma^z_{i,2}\sigma^z_{i,4}\right)
\right)
\end{eqnarray}

There are two triangular plaquettes in each unit cell and a conserved 
$Z_2$ flux associated with each of them. The flux operators are,
\begin{equation}
\label{lrflux}
W^L_i=\sigma^x_{i,1}\sigma^z_{i,2}\sigma^y_{i,3}~~~~
W^R_i=\sigma^x_{i,4}\sigma^z_{i,3}\sigma^y_{i,2}
\end{equation}
As in Kitaev's honeycomb model, these quantities are conserved as a 
consequence of a local spin rotation symmetry of the model. Namely, a 
$\pi$ rotation on each site of a plaquette about the direction of the 
outgoing bond.

Apart from these local conserved quantities there are also three global
qauntities which are conserved as a consequence the fact that a global
$\pi$ rotation about each of the three axes is a symmetry of the model.
We denote these by,
\begin{equation}
\label{Sigdef}
\Sigma^a\equiv e^{i\frac{\pi}{2}\sum_{i,a}\sigma^a_{i,\alpha}}
\end{equation}
It can be verified that $\Sigma^x$ is the product of the fluxes of all
the plaquettes,
\begin{equation}
\label{sigx}
\Sigma^x=\prod_i\left(W^L_iW^R_i\right)
\end{equation}

\section{Fermionisation}
\label{jwt}

We express the hamiltonian in terms of Majorana fermions using the
Jordan-Wigner transformation. We choose the Jordan-Wigner path to go
along the $x$ and the $y$ bonds from left to right. At every site we
have two bonds that are tangential to the path. We denote the incoming 
bond by $t_1$. This is the $x$-bond on the sublatttices 1, 3 and
$y$-bond on sublattices 2,4. The outgoing bond, corresponding to $y$ on 
sublattices 1,3 and $x$ on sublattices 2,4, is denoted by $t_2$. 
The third bond on each site which in normal to the path is denoted by
$n$ with the sign defined by $\hat n=\hat t_1\times\hat t_2$. With our
choice of path, the normal bond is $z$ for sublattices 1,3 and $-z$ for
sublattices 2,4. 

With these definitions the two Majorana fermions at each site are
defined as,
\begin{eqnarray}
\label{xidef}
\xi_{i,\alpha}&=&\sigma^{t_1}_{i,\alpha}
\prod_{j<i}\left(\prod_{\beta<\alpha}\sigma^n_{j,\beta}\right)\\
\label{etadef}
\eta_{i,\alpha}&=&\sigma^{t_2}_{i,\alpha}
\prod_{j<i}\left(\prod_{\beta<\alpha}\sigma^n_{j,\beta}\right)
\end{eqnarray}
The hamiltonian can be expressed in terms of the Majorana fermions
defined above,
\begin{eqnarray}
\nonumber
H&=&\sum_i\left(
J_x\left(i\xi_{i-1,4}\xi_{i,1}+i\xi_{i,2}\xi_{i,3}\right)\right.\\
\nonumber
&&+J_y\left(i\xi_{i,1}\xi_{i,2}+i\xi_{i,3}\xi_{i,4}\right)\\
\label{ham2}
&&\left.+J_z\left(-iu^L_i\xi_{i,1}\xi_{i,3}+iu^R_i\xi_{i,2}\xi_{i,4}\right)
\right)\\
\end{eqnarray}
where the link fields, $u^{L(R)}_i$ are defined as,
\begin{equation}
\label{urldef}
u^L_i\equiv i\eta_{i,3}\eta_{i,1}~~~~~~u^R_i\equiv i\eta_{i,2}\eta_{i,4}
\end{equation}
It is easy to see that the link fields are conserved quantities. Thus,
as expected for a generalised Kitaev model, the theory 
gets written in terms of Majorana fermions with nearest neighbour hopping 
in the background of conserved $Z_2$ gauge fields with the gauge fixing 
condition that the gauge fields on the $x$ and $y$ bonds are equal to $+1$.
It is easy to check that in this gauge the two flux operators are
proportional to the link fields,
\begin{equation}
\label{urlwrl}
u^L_i\equiv i\eta_{i,3}\eta_{i,1}=W^L_i
~~~~~~u^R_i\equiv i\eta_{i,2}\eta_{i,4}=W^R_i
\end{equation}

It is instructive to express the three global conserved quantities, 
$\Sigma^a$ in terms of the fermionic variables. We have,
\begin{eqnarray}
\label{fsigx}
\Sigma^x&=&\prod_i\left(\eta_{i,1}\eta_{i,2}\eta_{i,3}\eta_{i,4}\right)\\
\label{fsigy}
\Sigma^y&=&\prod_i\left(\xi_{i,1}\xi_{i,2}\xi_{i,3}\xi_{i,4}\right)\\
\label{fsigz}
\Sigma^z&=&\Sigma^x\Sigma^y
\end{eqnarray}
We will refer to $\Sigma^x$ as the flux number and $\Sigma^y$ as the Majorana
number.

\section{Diagonalisation}
\label{diag}

The hamiltonian can be diagonalised in the standard way. We write the
eigenstates as direct products of states in the $\eta$ fermion sector, 
$\vert{\cal G}\rangle$, which we will refer to as the gauge sector and
states in the $\xi$ fermion sector, $\vert{\cal M}\rangle$,  which we 
call the matter sector. We choose the states in the gauge sector to be
the simultaneous eigenstates of the $Z_2$ flux operators, i.e
$\vert{\cal G}\rangle=\vert\{w^L_i,w^R_i\}\rangle$, where
\begin{equation}
\label{fluxbasis}
u^{L(R)}_i\vert\{w^L_i,w^R_i\}\rangle=w^{L(R)}_i\vert\{w^L_i,w^R_i\}\rangle
\end{equation}
We then have
\begin{equation}
\label{sep}
H\left[u^L_i,u^R_i\right]\vert{\cal M}\rangle
\vert\{w^L_i,w^R_i\}\rangle=
H\left[w^L_i,w^R_i\right]\vert{\cal M}\rangle
\vert\{w^L_i,w^R_i\}\rangle
\end{equation}
The problem reduces to finding the eigenstates of the quadratic
hamiltonan of the $\xi$ fermions in the background of the gauge field
configuration $\{w^L_i,w^R_i\}$. The normal modes are given by the
solution of the eigenvalue equation,
\begin{equation}
\label{nmeq}
\sum_{j\beta}A_{i\alpha,j\beta}\phi^n_{j\beta}=\epsilon_n\phi^n_{i\alpha}
\end{equation}
where $A=T+V$ is a purely imaginary anti-symmetric matrix,
\begin{eqnarray}
\label{tdef}
T_{i,j}&=&iJ_x\left(\begin{array}{cccc}
0&0&0&-\delta_{i-1,j}\\
0&0&0&0\\
0&0&0&0\\
\delta_{i+1,j}&0&0&0\end{array}\right)\\
\nonumber\\
V_{i,j}&=&i\delta_{i,j}\left(\begin{array}{cccc}
0&J_y&-J_zw^L_i&0\\
-J_y&0&J_x&J_zw^R_i\\
J_zw^L_i&-J_x&0&J_y\\
0&-J_zw^R_i&-J_y&0\end{array}\right)
\end{eqnarray}

The eigenvalues come in pairs, $\pm\epsilon_n$. So we can always choose
$\epsilon_n$ to be positive. The eigenvectors corresponding to 
the positive and negative eigenvalues are complex conjugates. Denoting
the real and imaginary parts of the eigenvectors as $\phi^{nR(I)}$, we expand
the $\xi$ fermions as,
\begin{equation}
\label{modes}
\xi_{i,\alpha}=\sum_n a_n\phi^{nR}_{i,\alpha}+b_n\phi^{nI}_{i,\alpha}
\end{equation}
$a_n$ and $b_n$ are also Majorana fermions. The diagonal form of the 
hamiltonian is then,
\begin{equation}
\label{hdiag}
H=\sum_n~\epsilon_n~ib_na_n
\end{equation}
The ground state energy is the one where all the operators $ib_na_n$ are
diagonal and equal to -1. The ground state energy is,
\begin{equation}
\label{gsen}
E_0=-\sum_n~\epsilon_n
\end{equation}

\subsection{Boundary Conditions}
\label{bc}

We will be analysing the system with periodic (PBC) and open boundary 
conditions (OBC). The latter case is straightforward. The fermionic 
hamiltonian is exactly of the form given in equation (\ref{ham2}) with
$i=1,..,N$, where $N$ is the number of unit cells. The fermionic modes have
to be solved with the boundary conditions
\begin{equation}
\label{obc}
\phi_{0,a}=\phi_{N+1,a}=0
\end{equation}

PBC is a little more subtle. As is standard in Jordan-Wigner transformations,
the term in the hamiltonian for the link $i=N$ to $i=1$ is,
\begin{equation}
\label{llink}
H_{N,0}=J_x\Sigma^z i\xi_{1,1}\xi_{N,4}
\end{equation}
$\Sigma^z$ is a conserved quantity and can hence be chosen to be diagonal.
The fermionic modes have to be solved with periodic boundary conditions
for states with $\Sigma^z= +1$ and with anti-periodic boundary conditions for 
states with $\Sigma^z=-1$.
Namely,
\begin{equation}
\label{apbc}
\phi_{N+1,\alpha}=p\phi_{1,\alpha}
\end{equation}
where $p=\pm 1$ is the eigenvalue of $\Sigma^z$.

\section{The numerical solution}
\label{num}

We have solved eigenvalue problem numerically for a chain with $N$ unit 
cells and open boundary conditions for $N\le 5$. Each unit cell has 4
possible flux configurations making a total of $4^N$. We computed the single
particle spectrum for each flux configuration and calculated the lowest
total energy state for the range of values  $J_x=1.0,~J_y=0.1-10,~J_z=0.1-10$ 
(in the steps of 0.2). 

We find that the ground state is $2^N$ fold degenerate for all values of 
the parameters. This denegeracy comes from the fact that the single particle 
spectrum depends only on the values of the product, $w_i\equiv w^L_iw^R_i$, and
not on their individual values. i.e the spectrum only sees the total
flux passing through each unit cell which is the product of the values of the 
fluxes passing through the two triangular plaquettes that make up the unit cell.
 
The ground states correspond to the translationally invariant, $w_i=1$, flux 
configurations. 

\section{The Degeneracy of the States}
\label{degen}

As mentioned above, we find that the spectrum in each sector is $2^N$ fold
degenerate. This degeneracy is due to the fact that the energy
eigenvalues depend only on the total flux in the unit cell, 
namely $w_i\equiv w^L_iw^R_i$ and not on the individual fluxes, 
$w^L_i$ and $w^R_i$. In the remaining part of this section, we will give an
analytic proof of this $2^N$ fold degeneracy of all the states.

This degeneracy is related to but not the same as the $4^N$ degeneracy
that occurs in the simple Kitaev chain, the $J_z=0$ limit of our model.
In this case the degeneracy is easy to understand. The gauge fields do not 
occur at all in the hamiltonian. Thus each state is $4^N$ degenerate 
corresponding to all the states in the gauge sector. The extra $z$-bond 
terms in our model lift this degeneracy only partially.

At $J_y=J_z$, the denegeracy can be understood in terms of a local 
symmetry. It consists of interchanging the spins at sublattice 2 and 3
in any unit cell and then performing a $\pi/2$ rotation about 
the $x$-axis on all the spins in that unit cell. The operator that
implements this transformation is,
\begin{equation}
\label{pdef}
P_i\equiv\left(\frac{\vec\sigma_{i2}\cdot\vec\sigma_{i3}+1}{2}\right)
e^{i\frac{\pi}{4}\sum_{\alpha=1}^4\sigma^x_{i\alpha}}
\end{equation}
It is easy to verify that the $P_i$'s commute with the hamiltonian,
However $P_i$ flips the sign of the two flux operators in the unit cell,
\begin{equation}
\label{piprop2}
P_iW^{L(R)}_iP_i=-W^{L(R)}_i,~~~P_i^2=W^L_iW^R_i
\end{equation}
Thus it changes the flux configuration while conserving the total flux 
through the unit cell . Since it also does not change the energy
eigenvalue, it follows that every eigenstate of the hamiltonian is 
$2^N$ fold degenerate.

However, we numerically observe that the degeneracy persists even when
$J_y\ne J_z$. We will now give a proof for the degeneracy which is valid
at all couplings. We note that in equation (\ref{nmeq}), $\phi_{i,2}$ 
and $\phi_{i,3}$ couple only to sites within the unit cell. We express them 
in terms of $\phi_{i,1}$ and $\phi_{i,4}$ and obtain an eigenvalue equation 
for these quantities. We are then able to show that the eigenvalues
depend only on $w_i$.

We define the two component column vectors,
\begin{equation}
\label{chipsidef}
\chi_i\equiv\left(\begin{array}{c}\phi_{i,1}\\\phi_{i,4}\end{array}\right)
~~~~~~
\psi_i\equiv\left(\begin{array}{c}\phi_{i,2}\\\phi_{i,3}\end{array}\right)
\end{equation}
and the matrices,
\begin{eqnarray}
\label{t2def}
T_{ij}&=&i\delta_{i-1,j}\left(\begin{array}{cc}0&-J_x\\0&0\end{array}\right)
+i\delta_{i+1,j}\left(\begin{array}{cc}0&0\\ J_x&0\end{array}\right)\\
U_{ij}&=&i\delta_{i,j}\left(\begin{array}{cc}
J_y&J_zw^L_i\\ J_zw^R_i&-J_y\end{array}\right)
\end{eqnarray}
The eigenvalue equations (\ref{nmeq}) can be written as,
\begin{eqnarray}
\label{nmmat1}
\left(\begin{array}{cc}T&U\\ U^\dagger&J_x\tau^2\end{array}\right)
\left(\begin{array}{c}\chi\\ \psi\end{array}\right)
&=&\epsilon\left(\begin{array}{c}\chi\\ \psi\end{array}\right)
\end{eqnarray}
where $\tau^a,~a=1,2,3$ are the Pauli matrices.
$\psi$ can be eliminated from the equations to get,
\begin{equation}
\label{nmmat2}
\left(T+U\frac{1}{\epsilon-J_x\tau^2}U^\dagger\right)\chi
=\epsilon\chi
\end{equation}
The equations can be explicitly written as,
\begin{eqnarray}
\label{chieq2}
-iJ_x\phi_{i-1,4}+c_ie^{i\alpha_i}\phi_{i,4}&=&\lambda\phi_{i,1}\\
\label{chieq3}
iJ_x\phi_{i+1,1}+c_ie^{-i\alpha_i}\phi_{i,1}&=&\lambda\phi_{i,4}
\end{eqnarray}
where,
\begin{eqnarray}
\label{cdef}
c_i&=&\sqrt{\frac{J_x^2(J_y^2+J_z^2)^2+2J_z^2J_y^2(\epsilon^2-J_x^2)(1-w_i)}
{(\epsilon^2-J_x^2)^2}}\\
\label{alphadef}
\alpha_i&=&\tan^{-1}\left(\frac{\epsilon J_yJ_z}{J_x(J_y^2+J_z^2w_i)}
\left(w^R_i-w^L_i\right)\right)\\
\label{lamdef}
\lambda&=&\epsilon\left(1-\frac{J_y^2+J_z^2}{\epsilon^2-J_x^2}\right)
\end{eqnarray}
When these equations are solved for $\lambda$, they will yield an
equation for $\epsilon$. We now make a transformation,
\begin{eqnarray}
\label{gtrans1}
\phi_{i,1}&\rightarrow& e^{i\theta_i}\phi_{i,1}\\
\label{gtrans4}
\phi_{i,4}&\rightarrow& e^{i\theta_{i+1}}\phi_{i,4}\\
\label{thetadef}
\theta_i&=&-\sum_{j<i}\left(\alpha_j+\frac{\pi}{2}\right)
\end{eqnarray}
This gets rid of the phases in equations (\ref{chieq2}) and
(\ref{chieq3}) which become,
\begin{eqnarray}
\label{chieq4}
-iJ_x\phi_{i-1,4}-ic_i\phi_{i,4}&=&\lambda\phi_{i,1}\\
\label{chieq5}
iJ_x\phi_{i+1,1}+ic_i\phi_{i,1}&=&\lambda\phi_{i,4}
\end{eqnarray}
Since $c_i$ depends only on $w_i$ and not on $w^L_i$
and $w^R_i$ individually, $\lambda$ and hence $\epsilon$ depends only on
the $w_i$. Thus the energy eigenvalues depend only on the
total flux passing through the unit cell. 

This result is true for all values of $J_x,J_y$ and $J_z$. Note that when
$J_z=0$, equation (\ref{cdef}) implies that $c_i$ is independent of $w_i$
also. Thus the $4^N$ fold degeneracy of the simple Kitaev chain is recovered
in this limit.

\section{The Ground States and gaps}
\label{gs}

As we have mentioned earlier our numerical results show that the 
translationally invariant fluxes through the unit cells, namely $w_i=1$ is 
one of the ground states of the model. 
The fermionic problem is easy to solve analytically for these flux 
configurations and therefore the energy of the $w=1$ configuration can be 
compared to that of the $w=-1$ configuration in the thermodynamic limit. 

If $w_i=w$, then $c_i$ in equations (\ref{chieq4}, \ref{chieq5}) are 
independent of $i$, $c_i=c$. The equations can be solved by Fourier 
transforms,
\begin{eqnarray}
\phi_{i,a}&=&\int_{-\pi}^\pi\frac{dk}{2\pi}~e^{iki}\phi_a(k)\\
\phi_a(k)&=&\sum_i~e^{-iki}\phi_{i,a}
\end{eqnarray}
$\lambda$ is then given by the eigenvalues of the following matrix,
\begin{equation}
\label{kmat}
\left(\begin{array}{cc}0&-i\left(c+J_xe^{-ik}\right)\\
i\left(c+J_xe^{-ik}\right)&0\end{array}\right)
\end{equation}
namely,
\begin{equation}
\label{lamans}
\lambda=\pm\sqrt{c^2+J_x^2+2cJ_x\cos k}
\end{equation}
Along with equation (\ref{lamdef}) this yields an 
equation for the four energy bands. We have solved for this energy bands and
have computed the gap for $0<J_y,J_z<10$ and $J_x=1$. 
 
\begin{figure}[h] 
\includegraphics[width=\linewidth]{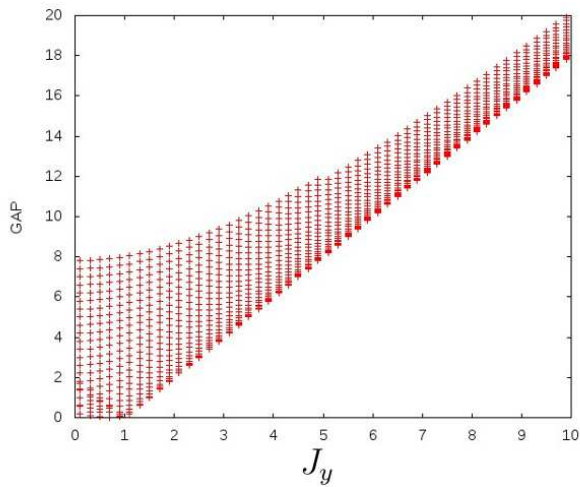}
\caption{ The gap plotted against $J_y$ for $J_x=1$ and values of $J_z$
varying in the range $0<J_z<10$. It can be seen that the $J_y<1$ region is 
gapless for some value of $J_z$.}
\label{gapvsjy}
\end{figure}

\begin{figure}[h] 
\includegraphics[width=\linewidth]{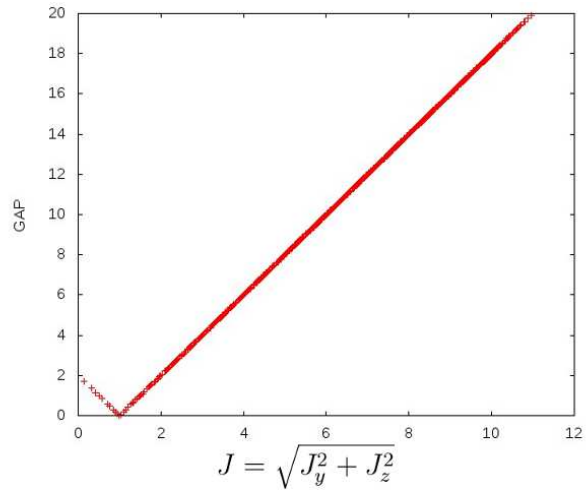}
\caption{ The gap plotted against $J\equiv\sqrt{J_y^2+J_z^2}$. The points 
fall on straight lines with slopes $\pm 2$}
\label{gapvsj}
\end{figure}

The calculations confirm that the $w_i=+1$ sector has a lower ground state 
energy than the $w_i=-1$ sector. The fermionic gap is twice the value of the
lowest single particle energy eigenvalue. This is shown in Figure 
\ref{gapvsjy} where the gap is plotted against $J_y$ for different 
values of $J_z$. The $J_y<1$ region is clearly gapless.
It is interesting that when we plot the gap as a function of 
$J\equiv\sqrt{J_y^2+J_z^2}$, all the points fall on the straight line as shown
in Fig. \ref{gapvsj}. This can be shown analytically since the expression
in equation (\ref{cdef}) simplifies considerably at $w_i=1$. It is then easy 
to show that the energy eigenvalues are given by,
\begin{equation}
\label{w1eev}
\epsilon^2=J^2+1\pm2J\cos\left(\frac{k}{2}\right)
\end{equation}
The gap is thus given by,
\begin{equation}
\label{lingap}
\Delta=2J_x\vert J-1\vert
\end{equation} 
The circle in the parameter space, $J_x^2=J_y^2+J_z^2$ is therefore gapless.

\section{Zero modes and Kinks}
\label{zmk}

We will now analyse the zero energy modes of the Majorana fermions. When
$\epsilon=0$, we have $\lambda=0$ and hence the equations (\ref{chieq4})
and (\ref{chieq5}) decouple and become very simple,
\begin{eqnarray}
\label{zm1}
J_x\phi_{i+1,1}+\frac{J_y^2+w_iJ_z^2}{J_x}\phi_{i,1}&=&0\\
\label{zm4}
J_x\phi_{i-1,4}+\frac{J_y^2+w_iJ_z^2}{J_x}\phi_{i,4}&=&0
\end{eqnarray}
These recursion relations can be formally solved,
\begin{eqnarray}
\label{phi1sol}
\phi_{i,1}&=&\prod_{j<i}\left(\frac{J_y^2+w_jJ_z^2}{J_x^2}\right)\phi_{1}\\
\label{phi4sol}
\phi_{i,4}&=&\prod_{j>i}\left(\frac{J_y^2+w_jJ_z^2}{J_x^2}\right)\phi_{4}
\end{eqnarray}
Where $\phi_{1(4)}$ are arbitrary constants. Thus, there are two formal 
solutions for every set of values of the parameters and every flux 
configuration. One with $\phi_{i,1}\ne 0$ and $\phi_{i,4}=0$ which
we denote by $\phi^+$ and the other with $\phi_{i,1}= 0$ and 
$\phi_{i,4}\ne 0$ which we denote by $\phi^-$.

However, the boundary conditions that the modes have to satisfy will pick out 
certain flux configurations for each point in the parameter space. We will 
analyse the situation for the cases of periodic boundary conditions (PBC) and 
open boundary conditions (OBC).

\subsection{Periodic Boundary Conditions}
\label{spbc}

We consider a chain with $N$ unit cells. As discussed in equation (\ref{apbc}),
PBC will imply that,
\begin{equation}
\label{pbc}
\phi_{N+1,a}=p\phi_{1,a}
\end{equation}

Equations (\ref{phi1sol}), (\ref{phi4sol}) and (\ref{pbc}) imply
\begin{equation}
\label{pbccond1}
\prod_{i=1}^N\left(\frac{J_y^2+w_jJ_z^2}{J_x^2}\right)=p
\end{equation}
Consider the general case where $M\le N$ of the $w_i$'s are equal to $-1$ and
$N-M$ of them are $+1$. We will refer to such configurations as $M$-defect
configurations. Equation (\ref{pbccond1}) gets written as,
\begin{equation}
\label{pbccond2}
\left(\frac{J_y^2-J_z^2}{J_y^2+J_z^2}\right)^M=p
\left(\frac{J_x^2}{J_y^2+J_z^2}\right)^N
\end{equation}
When $M=0$, we have only have solutions on the circle of radius $J_x$ in the 
$J_y-J_z$ plane. Note that $M=0$ is the translationally invariant ground state 
flux configuration. Thus this result implies that the model is gapless only on 
the circle, consistent with equation (\ref{lingap}).

When $M>0$, since the LHS of equation (\ref{pbccond2}) is less than $1$, 
no zero modes exist within the circle of radius $J_x$. Outside this circle, 
for every even $M$ there are two directions where equation (\ref{pbccond2}) 
is satisfied with $p=+1$. One direction where $J_y>J_z$ and the other where 
$J_z>J_y$. For odd $M$, the $J_y>J_z$ solution exists for $p=+1$ and the 
$J_y<J_z$ solution for $p=-1$.

\begin{figure}[h] 
\includegraphics[width=\linewidth]{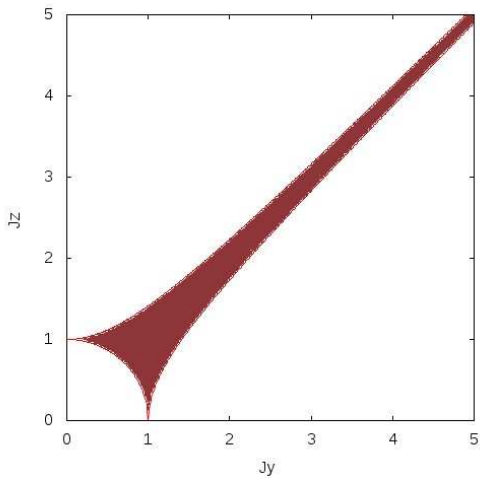}
\caption{ The region in the $J_y-J_z$ plane at $J_x=1$ that supports zero
energy modes}
\label{zmr}
\end{figure}

Thus for every $N$ there are a discrete set of points outside the circle 
which support zero energy modes. In the thermodynamic limit of 
$N\rightarrow\infty$, $N/M$ can take all values from $1$ to $\infty$. 
In this limit, all the points outside the circle in the range,
$J_x^2\le\vert J_y^2-J_z^2\vert\le 0$ support zero energy modes. This region
is shown for $J_x=1$ in Fig. \ref{zmr}

Consider the cases when the $M$ defects are in adjoining unit cells, say
from $i=1$ to $i=M$. We will call this a kink-antikink configuration.
We define polar coordinates in the $J_y-J_z$ plane,
\begin{equation}
J\equiv\sqrt{J_y^2+J_z^2},~~\gamma=\tan^{-1}\left(\frac{J_z}{J_y}\right)
\end{equation}
The unnormalised wave functions of the two zero modes are,
\begin{eqnarray}
\phi^{+}_{i,1}&=&J^{2(i-1)}~~~i\le M+1\\
&=&J{2(i-1)}\left(\cos2\gamma\right)^{i+1-M}~~~i>M+1\\
\phi^{+}_{i,4}&=&0\\
\phi^{-}_{i,4}&=&J^{-2(i-1)}~~~i\le M+1\\
&=&J^{-2(i-1)}\left(\cos2\gamma\right)^{-(i+1-M)}~~~i>M+1\\
\phi^{-}_{i,1}&=&0
\end{eqnarray}
$\phi^{\pm}_{i,2}$ and $\phi^{\pm}_{i,3}$ are given in terms of 
$\phi^{\pm}_{i,1}$ and $\phi^{\pm}_{i,4}$,
\begin{eqnarray}
\phi^{\pm}_{i,2}&=&
\frac{1}{J_x}\left(J_y\phi^\pm_{i,1}-w_i^RJ_z\phi^\pm_{i,4}\right)\\
\phi^{\pm}_{i,3}&=&
\frac{1}{J_x}\left(w_i^LJ_z\phi^\pm_{i,1}+J_y\phi^\pm_{i,4}\right)
\end{eqnarray}

It can be seen that $\phi^+$ is peaked at $i=M$ and is minimum at $i=1$ whereas
$\phi^-$ is peaked at $i=1$ and has a minmum at $i=M$. Thus we have one
Majorana mode localised at the location of the kink and another at the location
of the antikink. When $M$ is large these are well separated. Thus the situation
is similar to Kitaev's honeycomb model with kinks and antikinks playing the
role of the vortices. If the flux configuration can be manipulated, then so can
the Majorana modes trapped to them.

\subsection{Open Boundary Conditions}

\begin{figure}[h] 
\includegraphics[width=\linewidth]{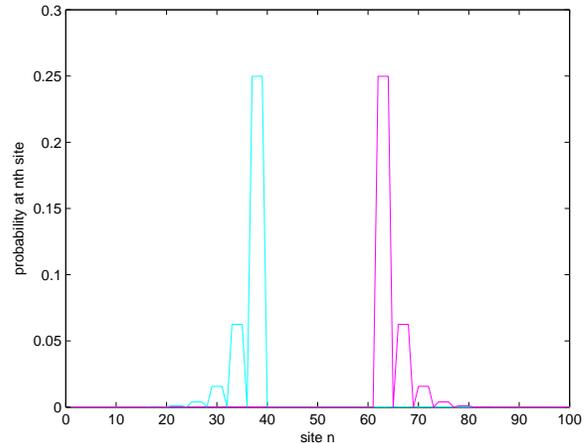}
\caption{The wavefunctions of the two Majorana zero modes for $N=25,~M=7$
with open boundary conditions.}
\label{7defect}
\end{figure}

We now consider open chains with $N$ unit cells. We then need to solve 
equations (\ref{zm1}) and (\ref{zm4}) with the boundary conditions
in equation (\ref{obc}).
From the solutions in equation (\ref{phi1sol}) and (\ref{phi4sol}), we see
that the above boundary conditions have non trivial solutions if and only
if at least one of the factors in the products on the RHS of the equations is
zero. This is only possible when $\vert J_y\vert=\vert J_z\vert$. Thus zero
modes exist in this case only at $\gamma=\pi/4$. At these points, the zero
modes are similar the ones in the PBC case except that the wavefunctions
strictly vanish in the region between the kink and the antikink. Namely,
if we consider a flux configuration with $w_i=-1,~i_i\le i<i_i+M$, then
the wavefunctions are given by,
\begin{eqnarray}
\phi^{+}_{i,1}&=&J^{2(i-1)}~~~i\le i_1\\
&=&0~~~i>i_1\\
\phi^{+}_{i,4}&=&0\\
\phi^{-}_{i,4}&=&J^{-2(i-1)}~~~i>i_1+M\\
&=&0~~~i>M+1\\
\phi^{-}_{i,1}&=&0
\end{eqnarray}
$\phi^{\pm}_{i,2}$ and $\phi^{\pm}_{i,3}$ are given in terms of 
$\phi^{\pm}_{i,1}$ and $\phi^{\pm}_{i,4}$ as before.
These wavefunctions are shown in Fig. \ref{7defect} for $N=25$ and $M=7$.

\subsection{Inhomogenous chains}

The solutions for the zero mode equations (\ref{zm1}) and (\ref{zm4}) hold
even for the case of inhomogenous chains where the coupling constants $J_y$
and $J_z$ depend on $i$. The equations then read
\begin{eqnarray}
\label{zm1ih}
J_{x}\phi_{i+1,1}+\frac{J_{yi}^2+w_iJ_{zi}^2}{J_x}\phi_{i,1}&=&0\\
\label{zm4ih}
J_x\phi_{i-1,4}+\frac{J_{yi}^2+w_iJ_{zi}^2}{J_x}\phi_{i,4}&=&0
\end{eqnarray}
These solution to these recursion relations is exactly the same as the case
of the homogenous chain with $i$ dependent $J_{y(z)}$.
\begin{eqnarray}
\label{phi1solih}
\phi_{i,1}&=&\prod_{j<i}\left(\frac{J_{yj}^2+w_jJ_{zj}^2}{J_x^2}\right)
\phi_{1}\\
\label{phi4solih}
\phi_{i,4}&=&\prod_{j>i}\left(\frac{J_{yj}^2+w_jJ_{zj}^2}{J_x^2}\right)
\phi_{4}
\end{eqnarray}
It is clear that by tuning the values of the site dependent couplings, a
large variety of zero mode wavefunctions can be engineered.

\subsection{Tuning the flux configuration}

We can add the following ``chemical potential" term for the conserved 
flux operators,
\begin{equation}
\label{chempot}
H_\mu=\sum_i~\left(\mu_i^LW_i^L+\mu_i^RW_i^R\right)
\end{equation}
This term will not change the eigenstates but will alter the energy 
eigenvalues. If the $\mu_i^{L(R)}$ can be tuned, then any particular
flux configurations can be made the ground state.

However it is still not known how to engineer these 3-spin operators 
in the physical realisations of the model in cold atom systems or in quantum 
circuits.

\section{Conclusions}
\label{concl}

To summarize, we have analysed an exactly solvable spin-1/2 chain which we call
a tetrahedral chain. This model is a generalization of Kitaev's honeycomb
lattice model. Like the honeycomb model, this model too has three coupling 
constants, $J_x$, $J_y$ and $J_z$. One of them can be scaled away, so without
loss of generality, we can set $J_x=1$. 

The model has conserved $Z_2$ fluxes on every triangular plaquette, namely
two fluxes per unit cell. We denote the values of these fluxes by $w^{L(R)}_i$.
The fermionic spectrum in the background of the flux configuration depends only
on the total flux in the unit cell, $w_i\equiv w^L_iw^R_i$. Consequently, every
eigenstate of the model is $2^N$ fold degenerate, where $N$ is the number of 
unit cells. 

The ground state flux configurations of the model are the ones with
$w_i=1$. The model is gapless on the unit circle in the $J_y-J_x$ plane 
and is gapped everywhere else. 

We have found simple analytic solutions for zero energy wavefunctions for the 
Majorana fermions. We have delineated the region in parameter space where 
these solutions exist. In the uniform case, the two zero modes are peaked
at the kink and anti-kink positions and are well separated if the kink and
antikink are well separated. The analytic solutions also show how the 
zero mode wavefunctions can be engineered by tuning the couplings $J_y$ and 
$J_z$ in a site dependent way.

In conclusion our results show how zero energy Majorana modes can be created
and manipulated in the tetrahedral spin chain if it is possible to tune the
flux configurations and the local couplings. In this work, we have not
addressed the interesting and important question as to how this is useful to
braid unpaired Majorana modes. While this is of course not a meaningful
operation in a chain, it may be possible to implement braiding in a network of
coupled chains using the results obtained in this paper. We are pursuing this
line of work and hope to report on it in future.

\section*{Acknowledgments}

We are grateful to G. Baskaran for motivating this work and for
useful discussions.


\begin{thebibliography}{99}

\bibitem{tqc}
C. Nayak, S. H. Simon, A. Stern, M. Freedman and S. D. Sarma, Rev. Mod. Phys. \textbf{80}, 1083 (2008).
\bibitem{umf}
Sumanta Tewari, S. Das Sarma, Chetan Nayak, Chuanwei Zhang and P. Zoller, Phys. Rev. Lett. \textbf{98}, 010506 (2007).
\bibitem{mooreread}
G. Moore and N. Read, Nucl. Phys. B \textbf{360, 362} (1991).
\bibitem{faroaetal}
Lara Faoro, Jens Siewert and Rosario Fazio, Phys. Rev. Lett. \textbf{90}, 028301 (2003).
\bibitem{kitaevhc}
A. Y. Kitaev, Ann. Phys. (N.Y) \textbf{303}, 2 (2003), A. Y. Kitaev, Ann. Phys. (N.Y) \textbf{321}, 2 (2006).
\bibitem{gbpwa}
G. Baskaran, Z. Zou and P. W. Anderson, Solid State Commun. \textbf{63}, 973 (1987). 
\bibitem{demler}
L. M. Duan, E. Demler and M. D. Lukin, Phys. Rev. Lett. \textbf{91}, 090402 (2003).
\bibitem{nori}
J. Q. You, Xiao- Feng Shi, Xuedong Hu and Franco Nori, 
Phys. Rev. B \textbf{81}, 014505 (2010). 
\bibitem{1dkit}
A. Kitaev, Usp. Fiz. Nauk (Suppl.) \textbf{171}, 131 (2001).
\bibitem{1dumf}
Uma Divakaran and Amit Dutta, Phys. Rev. B \textbf{79}, 224408 (2009).
\end{thebibliography}
\end{document}